%% file: 0_main.tex
\documentclass[manuscript=article, journal=mamobx, layout=twocolumn]{achemso}
\setkeys{acs}{doi = false}
\renewcommand\tocentryname{}

\makeatletter
\renewcommand*{\acs@tocentry@print}[1]{%
    \gdef\acs@tocentry@text{\normalsize#1}%
    \acs@tocentry@print@aux%
}
\renewcommand*{\acs@tocentry@print@aux}{%
    \begingroup 
    \let\@startsection\acs@startsection@orig 
    \acs@section*{\tocentryname}%
    \centering
    {\textbf{For Table of Contents}}\hfill
    \tocsize 
    \sffamily 
    \singlespacing 
    \begin{center} 
            {%
                \begin{minipage}{\acs@tocentry@width} 
                \vbox to \acs@tocentry@height{\acs@tocentry@text}%
                \end{minipage}%
            }%
    \end{center}%
    \endgroup 
}
\makeatother

\let\oldmaketitle\maketitle
\let\maketitle\relax

\usepackage{chemformula}
\usepackage{subcaption}

\newcommand{\angstrom}{\textup{\AA}}
\newcommand*{\br}[1][]{\mathbf{r}_{#1}}                  
\newcommand*{\bR}{\mathbf{R}}                            
\newcommand*{\dr}[1][]{\mathrm{d}\mathbf{r}_{#1}\ }      
\newcommand*{\drsup}[1][]{\mathrm{d}\mathbf{r}^{#1}\ }   
\newcommand*{\dvar}[2][]{\mathrm{d #2}_{#1}\ }           

\author{Takashi J. Yokokura}
\affiliation[University of California Berkeley]
{Department of Chemical and Biomolecular Engineering, University of California Berkeley, Berkeley, California 94720, USA}
\author{Chao Duan}
\affiliation[University of California Berkeley]
{Department of Chemical and Biomolecular Engineering, University of California Berkeley, Berkeley, California 94720, USA}
\author{Rui Wang}
\email{ruiwang325@berkeley.edu}
\affiliation[University of California Berkeley]
{Department of Chemical and Biomolecular Engineering, University of California Berkeley, Berkeley, California 94720, USA}
\alsoaffiliation[Lawrence Berkeley National Lab]
{Materials Sciences Division, Lawrence Berkeley National Lab, Berkeley, California 94720, USA}

\title{Microphase Segregation in Polyelectrolyte Brushes}


\begin{document}

\twocolumn[
\begin{@twocolumnfalse}
\oldmaketitle
\begin{abstract}
Polyelectrolyte (PE) brushes have ubiquitous applications as surface modifiers which regulate various structural and dynamic properties.
Here, we apply a continuous-space self-consistent field theory to study the structural heterogeneity in PE brushes induced by competing electrostatic and hydrophobic interactions.
For brushes with high grafting densities, we find a series of microphase-segregated morphologies with alternating polymer-rich and polymer-poor layers in the direction normal to the substrate.
The transitions between multi-layer morphologies with consecutive numbers of condensed layers as well as the melting transition to the fully swollen brush are all discontinuous. 
We also elucidate that the segregated layers are formed by different subpopulations of all chains, significantly different from the scenario of the pearl-necklace structure formed by a single PE in poor solvents. 
Furthermore, we bridge the microstructure of multi-layer brushes to experimentally measurable reflectivity spectra, where the oscillation period and amplitude in the spectra are shown to be very sensitive to the number of layers and the sharpness of the polymer-solvent interface. 
The multi-layer morphology predicted by our theory is in good agreement with the lamellae structure experimentally observed at hydrated Nafion film interfaces. 
\nolinebreak
\begin{tocentry}
    \includegraphics{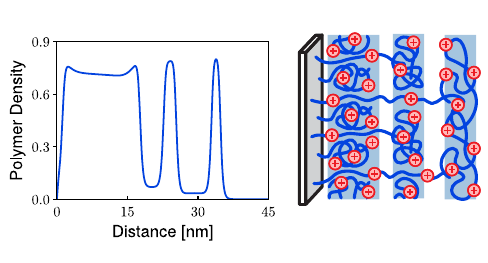}
\end{tocentry}

\end{abstract}
\end{@twocolumnfalse}
]

\input{1_introduction}
\input{2_modeltheory}
\input{3_resultsdiscussion}

\input{4_conclusions}

\begin{acknowledgement}
This material is based upon work supported by the National Science Foundation Graduate Research Fellowship under Grant No. DGE 2146752 to TJY. Acknowledgment is made to the donors of the American Chemical Society Petroleum Research Fund for partial support of this research. This research used the computational resource provided by the Kenneth S. Pitzer Center for Theoretical Chemistry.
\end{acknowledgement}

\begin{suppinfo}
Effect of bulk salt concentration on the brush morphology; electric double layer structure within multi-layer morphologies; effect of polymer dielectric constant on the stability of multilayer morphologies; effect of bulk salt concentration on the reflectivity spectra.
\end{suppinfo}

\bibliography{sysg_20240725}

\include{5_SI}

\end{document}

%% file: 1_introduction.tex
\clearpage
\section*{Introduction}
Polyelectrolyte (PE) brushes are a collection of charged polymers grafted onto a surface and can be designed with varying chain architectures, copolymer compositions and charge patterns \cite{chen_50th_2017, das_polyelectrolyte_2015}. PE brushes are widely used as surface modifiers to regulate surface properties like transport, wettability, adhesion, antifouling and lubrication \cite{xiong_neuromorphic_2023, xin_reversibly_2010, stuart_emerging_2010, he_tuning_2016}. 
Charged polymers are sensitive to external stimuli where their responses to temperature, mechanical deformation and ionic environment (e.g., concentration, valency and specific ion) depend on the chemical identities of their constituent monomers \cite{ruhe_polyelectrolyte_2004, conrad_towards_2019}. As a result, PE brushes are attractive candidates for stimuli-responsive smart materials and devices, such as valves \cite{ito_signal-responsive_2000, yameen_single_2009}, sensors \cite{wiarachai_clickable_2016, xie_chiral_2015, welch_polymer_2011}, actuators \cite{kelby_controlled_2011, ionov_reversible_2006} and drug delivery vehicles \cite{liechty_polymers_2010}. 
Furthermore, because biomacromolecules like proteins and RNA are essentially PEs, the study of PE brushes also provides fundamental understanding on the structure and functionality of biologically relevant systems, such as neurofilaments \cite{zhulina_self-consistent_2007}. 

In PE brushes, the interplay between electrostatic interactions due to their charged monomers and hydrophobic interactions due to their hydrocarbon chain backbones gives rise to rich morphological behaviors. Using surface force apparatus, Yu et al. found that the morphologies and mechanical responses of synthetic polystyrenesulfonate (PSS) brushes in multivalent salt solutions are quite non-trivial \cite{yu_multivalent_2018}. For instance, the brush heights change non-monotonically, collapse followed by re-expansion. It was reported that the friction force between opposing PE brushes increases dramatically with the addition of multivalent salt, diminishing the lubricity. 
Srinivasan et al. grafted recombinant intrinsically disordered proteins onto a substrate and observed a dramatic height change at a critical monovalent salt concentration \cite{srinivasan_stimuli-sensitive_2014}. They also found that brush height is heavily influenced by the charge distribution of the constituent proteins due to amino acid sequence and phosphorylation state \cite{bhagawati_site-specific_2016, lei_structural_2018}.

Interestingly, the competing interactions can induce brush morphologies beyond the homogeneous structure,  with segregated micro-domains instead. 
Based on atomic force microscopy (AFM) images, Yu et al. found that PE brushes form pinned micelles in multivalent salt solutions where chains aggregate laterally (i.e., in-plane to the grafting substrate) \cite{Yu_multivalent_2017}.
Using molecular dynamics simulations, Carrillo and Dobrynin reported that PE brushes in monovalent salt solutions exhibit morphologies containing cylindrical aggregates, driven by the hydrophobic attraction between different monomers \cite{carrillo_morphologies_2009}. Similar to pinned micelles, these cylindrical aggregates are also formed by different chains but are stretched away from the substrate. 
In addition to the lateral direction, it is worth noting that the structural heterogeneity can also occur in the vertical direction normal to the substrate. 
Samokhina et al. grafted PSS chains on spherical particles \cite{samokhina_binding_2007}. They found the coexistence between an inner condensed layer and an outer dilute layer via cryogenic transmission electron microscopy (cryo-TEM).  Similar morphologies with coexisting layers were also detected by neutron reflectivity \cite{reinhardt_fine-tuning_2013, yim_evidence_2005}. 
Furthermore, Dura et al. and Randall et al. reported lamellae composed of alternating layers of ionomer Nafion and water at the interface between the substrate and the hydrated Nafion film using neutron reflectivity \cite{dura_multilamellar_2009, randall_morphology_2024}. 
All the aforementioned brush morphologies with structural heterogeneity are significantly different from the scenario depicted by the well-known Alexander---de Gennes model. 

Uncovering the microstructure of PE brushes using state-of-the-art experimental techniques remains a great challenge. 
Computational methods provide an alternative solution \cite{de_gennes_conformations_1980, pincus_colloid_1991, israels_charged_1994, borisov_diagram_1994, zhulina_theory_1995, wijmans_self-consistent_1992, scheutjens_statistical_1979, scheutjens_statistical_1980, leermakers_modeling_2010, zhulina_effect_2007, jiang_structure_2007, qing_interfacial_2022, jiang_density_2018, li_density_2006, prusty_modeling_2024, morochnik_structural_2017, prusty_charge_2020, li_effects_2022, rumyantsev_surface-immobilized_2023, vigil_self-consistent_2022}. 
Great efforts have been made to theoretically model PE brushes since the pioneering work of de Gennes and Pincus \cite{de_gennes_conformations_1980, pincus_colloid_1991}. 
Built upon the Alexander---de Gennes model, Pincus developed a scaling theory which utilized a force balance between the repulsive osmotic pressure due to the translational entropy of ions and the elastic attraction due to chain deformation. 
Zhulina and coworkers presented a diagram of states including the osmotic, salted and neutral brush regimes \cite{israels_charged_1994, borisov_diagram_1994, zhulina_theory_1995}. 
In addition, lattice self-consistent field theory (commonly known as the Scheutjens---Fleer model) has been widely applied to describe synthetic and protein-inspired PE brushes \cite{wijmans_self-consistent_1992, scheutjens_statistical_1979, scheutjens_statistical_1980, leermakers_modeling_2010, zhulina_effect_2007}. However, the Scheutjens---Fleer model constrains each monomer to a lattice site, which limits its flexibility in describing complex heterogeneous morphologies. 
Furthermore, classical density functional theory (CDFT) has also been used to investigate PE brush morphologies. The predicted scaling relationship between the brush height and salt concentration is in agreement with scaling theory and simulations \cite{jiang_structure_2007, qing_interfacial_2022}. However, CDFT largely relies on the preassumed functional form of the electrostatic correlation and intra-chain correlation, where improper choices can lead to non-physical results \cite{jiang_density_2018, li_density_2006, prusty_modeling_2024}. 

The coexistence of local segregation and long-range stretching is an important conformational feature of PE\cite{muthukumar_50th_2017, muthukumar_physics_2023, duan_electrostatics-induced_2023, duan_association_2022}. This may lead to the emergence of microphase separation in a variety of PE systems \cite{joanny_weakly_1990, borue_statistical_1988, dobrynin_hydrophobic_1999}. 
A well-known example is the pearl-necklace structure formed by a single PE in poor solvents \cite{dobrynin_cascade_1996, dobrynin_theory_2005, duan_conformation_2020, duan_stable_2022}. 
Recently, Liu et al. developed a variational approach to control the center-of-mass of polymers \cite{liu_variational_2024}. Incorporating this method into the self-consistent field theory (SCFT), they are able to predict the pearl-necklace structure formed by a single PE. The predicted polymer conformation, scaling behavior and cascade transition is in good agreement with the classical work by Dobrynin, Rubinstein, and Obukhov \cite{dobrynin_cascade_1996}.
It is thus natural to ask how this structure changes when PEs are collectively tethered to a surface. 
The study of structural heterogeneity with local segregation is critical for understanding the morphology, surface properties and interactions of PE brushes in order to fully realize their potential as surface regulators. 
In an earlier work, Duan et al. predicted that structural heterogeneity is preferred in the lateral direction for brushes with lower grafting densities, whereas brushes with higher grafting densities favor heterogeneous morphologies in the normal direction \cite{duan_electrostatic_2024}. 
In this work, we apply continuous-space SCFT to systematically investigate the effects of the competing electrostatic and hydrophobic interactions on the structural heterogeneity of PE brushes with high grafting densities. 
We find cascade transitions between a series of microphase-segregated morphologies with alternating polymer-rich and polymer-poor layers in the direction normal to the substrates. 
The molecular picture underlying these multi-layer morphologies is also discussed. 
Furthermore, we examine the evolution of scattering spectra accompanying the morphological changes.

%% file: 2_modeltheory.tex
\section*{Model and Theory}
\begin{figure}
    \centering
    \includegraphics[width=0.50\linewidth]{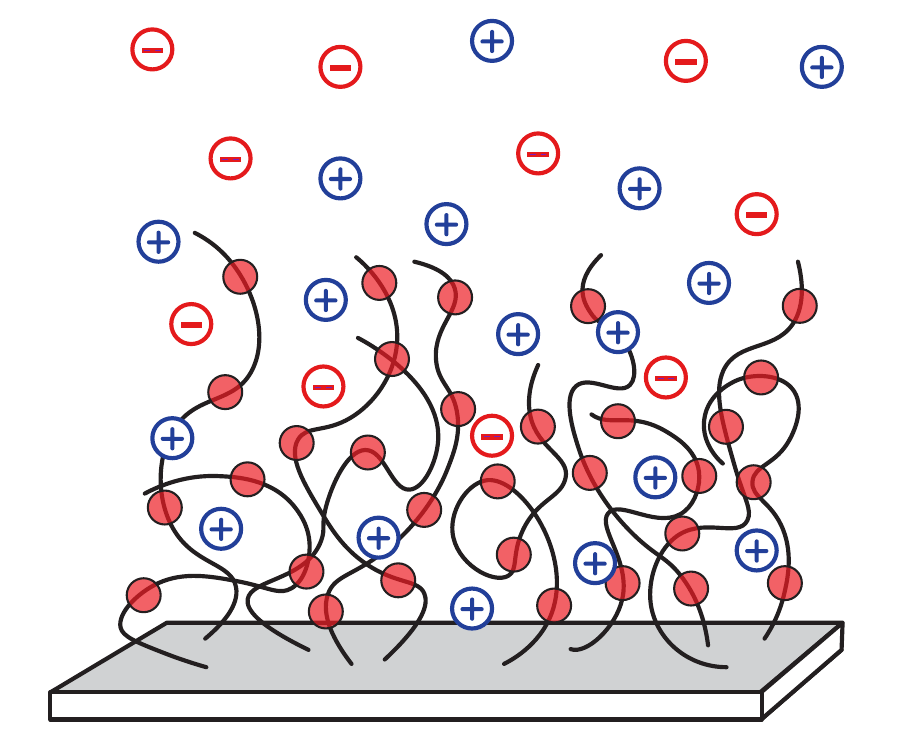}
    \caption{Schematic of a polyelectrolyte brush on a planar substrate immersed in a salt solution.}
    \label{fig:schem}
\end{figure}

As shown in Fig.~\ref{fig:schem}, we consider brushes composed of polyelectrolyte (PE) chains with one end grafted on a substrate and immersed in a solution of $n_s$ solvent molecules and $n_\pm$ mobile ions. The system is treated as a semicanonical ensemble with a fixed number of PE chains $n_p$ whereas solvent and mobile ions are connected to a bulk salt solution of ion concentration $c_\pm^b$ that maintains the chemical potentials of the solvent $\mu_s$ and ions $\mu_{\pm}$. PEs are assumed to be Gaussian chains of $N$ total segments with Kuhn length $b$. We adopt the smear charge model with the backbone charge density $\alpha$. Mobile ions are considered to have valency $z_\pm$. 

The semicanonical partition function of the system is
\begin{align}
\Xi =~&\frac{1}{n_p!\nu_p^{N n_p}} \prod_{\gamma} \sum_{n_\gamma=0}^{\infty}\frac{e^{\mu_\gamma n_\gamma}}{n_\gamma!v_\gamma^{n_\gamma}}\label{eqn:Xi}\\ 
~&\prod_{i=1}^{n_p} \int \mathcal{D}\{\bR_i\} \prod_{j = 1}^{n_\gamma} \int \dr[\gamma, j]  \exp \bigg\{-\mathcal{H} \bigg\}\nonumber\\ 
~&\prod_{\br} \delta[\hat{\phi}_p(\br) + \hat{\phi}_s(\br) -1]~,\nonumber 
\end{align}
where small molecules are denoted $\gamma=s, \pm$ for solvent and ions, respectively. $\nu_p$ is the volume of PE segments, whereas $\nu_\gamma$ is the volume of small molecules. For simplicity, we assume $\nu_p = \nu_s = \nu$. $\mathcal{D}\{\bR_i \}$ denotes the integration over all chain configurations for each chain $i$. $\hat{\phi}_p(\br)$ and $\hat{\phi}_s(\br)$ are the local instantaneous volume fractions of the PE and solvent, respectively. The $\delta$ function at the end of Eq.~\ref{eqn:Xi} accounts for the incompressibility. 
We only consider the contributions from polymers and solvent in the incompressibility constraint. For the parameter space studied in this work, the volume fractions of ions in the brushes are very low and can thus be neglected.

The Hamiltonian $\mathcal{H}$ is given by 
\begin{align}
\mathcal{H} =~&\sum_{k=1}^{n_p} \frac{3}{2b^2} \int_{0}^{N} \dvar{s} \bigg(\frac{\partial \bR_k(s)}{\partial s}\bigg)^2\label{eqn:H}\\ &+ 
\frac{1}{\nu} \int \dr \chi \hat{\phi}_p(\br)\hat{\phi}_s(\br) \nonumber\\
&+ \frac{1}{2} \int \dr \drsup[\prime]\ \hat{\rho}_c(\br)\ C(\br, \br^\prime)\ \hat{\rho}_c(\br^{\prime})~,\nonumber
\end{align}
which contains the elastic energy governed by Gaussian chain statistics, the short-range hydrophobic interactions between polymer segments and solvent molecules, and the long-range Coulomb interactions between all charged species. The hydrophobic interaction is manifested by the Flory---Huggins $\chi$ parameter. $\hat{\rho}_c(\br)  = z_+\hat{c}_+(\br) - z_-\hat{c}_-(\br) + \alpha \hat{\phi}_p(\br)/\nu$ is the local charge density with $\hat{c}_\pm(\br)$ as the instantaneous number density of ions. $C(\br, \br^{\prime})$ is the Coulomb operator satisfying $-\nabla\cdot\big[\epsilon(\br)\nabla C(\br, \br^{\prime})\big]=\delta(\br- \br{^\prime})$. $\epsilon(\br) = kT \epsilon_0\epsilon_r(\br)/e^{2}$ is the scaled permittivity, where  $\epsilon_0$ is the vacuum permittivity, $e$ is the elementary charge, and $\epsilon_r(\br)$ is the local dielectric constant dependent on the local composition of the system \cite{nakamura_salt-doped_2012, sing_electrostatic_2014, wang_effects_2011, hou_solvation_2018}. 

We follow the standard self-consistent field procedure \cite{fredrickson_equilibrium_2006} by first decoupling the interacting system into non-interacting PE chains and ions in fluctuating fields using the Hubbard---Stratonovich transformation and identity transforms. We then employ the saddle-point approximation to simplify the evaluation of the functional integral over the fluctuating fields. These give rise to the following self-consistent equations for polymer density $\phi_p(\br)$, conjugate fields $w_p(\br)$ and $w_s(\br)$, electrostatic potential $\psi(\br)$ and ion concentrations $c_\pm(\br)$:

\begin{subequations} \label{SCF:main}
\begin{align}
&w_p(\br) - w_s(\br) = \chi(1-\phi_p(\br))+\alpha\psi(\br)\label{SCF:a}\\ 
&\quad\quad\quad\quad\quad\quad\quad-\frac{\partial\epsilon(\br)}{\partial\phi_p}\nu\lvert \nabla \psi(\br) \rvert^2  -\chi\phi_p(\br)\nonumber\\ 
&\phi_p(\br) = ~\frac{n_p}{Q_p} \int_{0}^{N} \mathrm{ds}\ q(\br;\mathrm{s}) q_c(\br;\mathrm{s})\label{SCF:b}\\
&1-\phi_p(\br) = e^{\mu_s}\exp{(-w_s(\br))}\label{SCF:c}\\
&-\nabla\cdot \big(\epsilon(\br)\nabla\psi(\br) \big) =\label{SCF:d}\\ 
&\quad\quad\quad\quad\quad\quad\quad z_+c_+(\br)-z_-c_-(\br)+\frac{\alpha}{\nu} \phi_p(\br)~,\nonumber
\end{align}
\end{subequations}
where $c_\pm = \lambda_\pm\exp(\mp z_\pm\psi)$ is the ion concentration with $\lambda_\pm = e^{\mu_\pm}/\nu_\pm $ the fugacity of the ions determined by the bulk ion concentration $c^b_\pm$. $Q_p=\nu^{-1}\int \dr q(\br; \mathrm{s})$ is the single-chain partition function, where $q(\br; \mathrm{s})$ is the chain propagator satisfying the modified diffusion equation:
\begin{align}\label{eqn:MDE}
\frac{\partial}{\partial \mathrm{s}} q(\br;\mathrm{s}) = \frac{b^2}{6}&\nabla^2 q(\br;\mathrm{s})- w_p(\br)q(\br;\mathrm{s})~.
\end{align}
Here, we consider PE brushes grafted onto a planar substrate with sufficiently high grafting densities and chain lengths such that the polymer densities only vary in the direction normal to the substrate $z$ but remain homogeneous in the $xy$-plane. The initial condition for the propagator beginning at the surface is $q(\mathrm{z}; 0) = \delta(\mathrm{z}-\mathrm{z}*)$, where $\mathrm{z}*\to 0_+$ is the grafting plane. On the other hand, $q_c(\mathrm{z}; \mathrm{s})$ introduced in Eq.~\ref{SCF:b} is the complementary propagator that begins at the free end of the chain and also follows the modified diffusion equation (Eq.~\ref{eqn:MDE}) but with the initial condition $q_c(\mathrm{z}; N) = 1$. The resulting free energy per unit area in excess to the reference bulk salt solution is then:
\begin{align}
F =&-\sigma\ln Q_p - e^{\mu_s}Q_s+\frac{1}{\nu}\int \dvar{z} \Big(\chi\phi_p (1-\phi_p) \nonumber\\ 
-& w_p\phi_p - w_s(1-\phi_p) \Big)+ \int \dvar{z} \Big(\frac{\alpha}{\nu}\phi_p\psi \label{eqn:F}\\ -&\frac{\epsilon}{2}\bigg( \frac{\mathrm{d}\psi}{\mathrm{dz}} \bigg)^2 - c_+-c_-+c_+^b+c_-^b\Big)~.\nonumber
\end{align}
$\sigma = n_p/A$ is the grafting density of chains on the substrate, where $A$ is the total area of the substrate. $Q_s=\nu^{-1} \int \dvar{z} \exp(-w_s)$ is the solvent partition function.  
Here we consider the substrate to be impenetrable and charge neutral, which is reflected by setting the boundary conditions $\phi_p=0$ and $\mathrm{d}\psi/\mathrm{dz}=0$ at the surface ($z=0$). 

Eq.~\ref{SCF:main} is solved iteratively until self-consistency is achieved and the equilibrium PE brush density distribution, electrostatic potential, and ion distribution can be obtained. 
The modified diffusion equation (Eq.~\ref{eqn:MDE}) is solved by using the Crank---Nicolson algorithm whereas the Poisson---Boltzmann equation (Eq.~\ref{SCF:d}) is solved using an iterative, centered finite difference scheme \cite{hoffman_numerical_2018}. 
In contrast to the Scheutjens---Fleer (SF) model with the lattice constraint \cite{scheutjens_statistical_1979, scheutjens_statistical_1980}, the differential equations in our work are solved in the continuous space, which decouples the numerical discretization from the physical lattice. 
This improves both the accuracy and flexibility of the calculations, enabling us to explore morphologies with more complex heterogeneity \cite{chantawansri_spectral_2011}. 
Furthermore, our theory can not only be easily generalized to polyelectrolytes with various chain architectures, copolymer composition and  charge patterns, but also be straightfowardly implemented to biomacromolecules like disordered protein brushes \cite{yokokura_effects_2024, ding_dissecting_2024}. 
In the current version of our SCFT, the electrostatics is treated at the mean-field level. The ion--ion correlations can be included by involving the Gaussian fluctuation theory into the treatment of the electrostatic fields 
\cite{agrawal_electrostatic_2022, duan_electrostatic_2024}. We also note that, compared to methods like molecular dynamics simulations, our current theory is limited to modeling equilibrium morphologies and cannot capture the kinetics of their transitions and other dynamic properties.

%% file: 3_resultsdiscussion.tex
\begin{figure*}[htbp]
    \centering
    \includegraphics[width=0.8\textwidth]{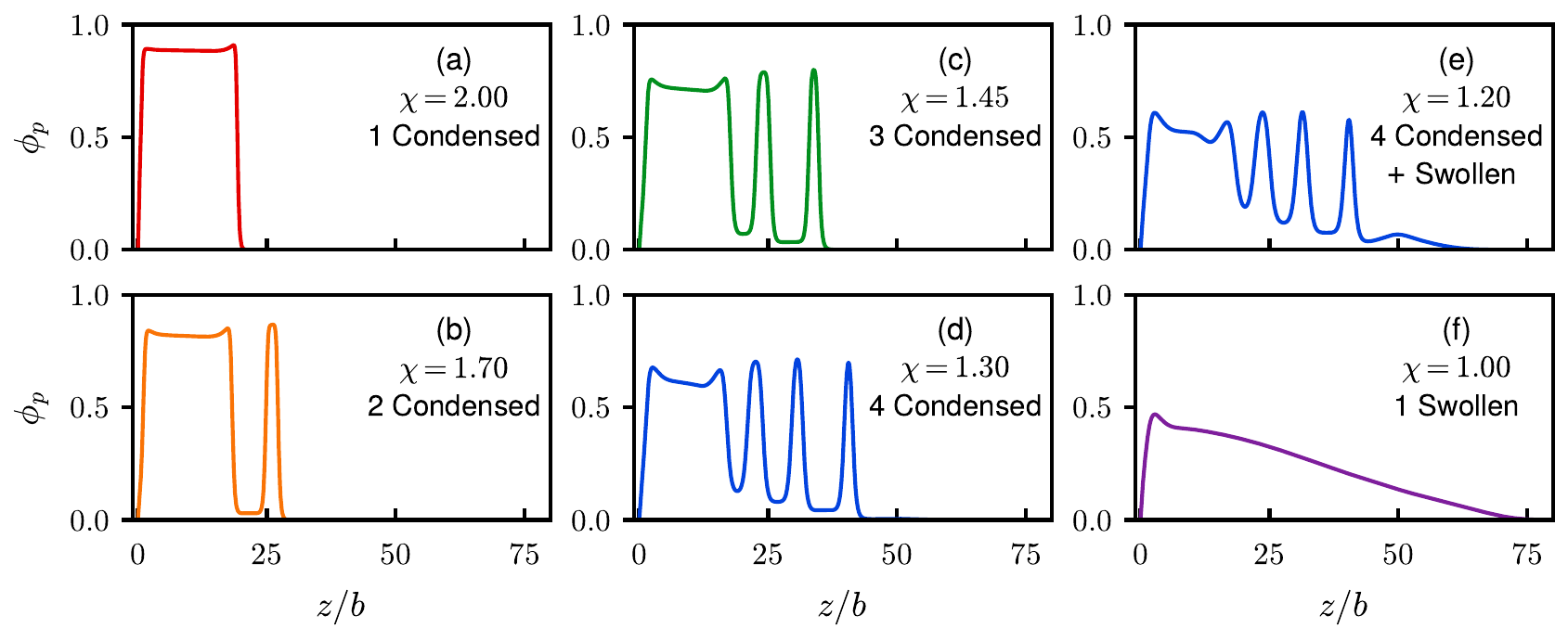} 
    \caption{Evolution of multi-layer morphologies in PE brushes. Polymer density distributions are plotted for varying hydrophobicities $\chi$ with fixed charge density $\alpha=0.4$: (a) 1 condensed layer at $\chi=2.00$; (b) 2 condensed layers at $\chi=1.70$; (c) 3 condensed layers at $\chi=1.45$; (d) 4 condensed layers at $\chi=1.30$; (e) 4 condensed layers with an additional, outer swollen layer at $\chi=1.20$; and (f) fully swollen brushes at $\chi=1.00$. The $x$-axis $z/b$ denotes the distance (scaled by the Kuhn length) from the grafting surface in the normal direction.}
    {\phantomsubcaption{}\label{fig:pha1C} \phantomsubcaption{}\label{fig:pha2C} \phantomsubcaption{}\label{fig:pha3C} \phantomsubcaption{}\label{fig:pha4C} \phantomsubcaption{}\label{fig:pha4Cd} \phantomsubcaption{}\label{fig:phaS}}
    \label{fig:pha}
\end{figure*}

\section*{Results and Discussion}

In this work, we focus on the competing interactions between electrostatic repulsion and hydrophobic attraction on the morphology of PE brushes. These two interactions are controlled by the backbone charge density $\alpha$ and the Flory---Huggins $\chi$ parameter, respectively. 
We consider PE brushes with grafting density $\sigma = 0.02\ \mathrm{nm}^{-2}$ immersed in a monovalent salt ($z_\pm = 1$) solution with bulk concentration $c_\pm^b = 10$ mM.
The temperature is set at 293 K, yielding a Bjerrum length $l_B = e^2/4\pi kT\epsilon_0 \epsilon_r$ of 0.7 nm, and a Debye screening length $\kappa_D^{-1} = [4 \pi l_B (z_+c_+^b + z_-c_-^b)]^{-1/2}$ of 3 nm. Each constituent polymer has a chain length of 200 segments with the Kuhn length $b = 1.0$ nm.
For this set of parameters, the brushes studied in this work are within the dense brush regime. For example, with $\alpha=0.4$, $R_g^2 \sigma \sim (\alpha N b)^2 \sigma = 128 \gg 1$, where $R_g$ is the radius of gyration of a polyelectrolyte.

\subsection*{Evolution of Multi-layer\\ Morphologies}

It is well-known that a single PE chain in poor solvents can form a series of pearl-necklace structures induced by the competition between long-range electrostatic repulsion and short-range hydrophobic attraction \cite{dobrynin_theory_2005, minko_single_2002}. 
Similar heterogeneous structures with local polymer segregation also exist in PE brushes. 
Fig.~\ref{fig:pha} shows that, as the hydrophobicity $\chi$ decreases with the charge density fixed at $\alpha=0.4$, a series of microphase-segregated multi-layer morphologies appear. 
When hydrophobic attraction is dominant ($\chi=2.0$), all the polymers are compressed into a single condensed layer to minimize the contact with the solvent. 
As $\chi$ decreases, hydrophobic attraction is reduced---equivalently, the relative importance of electrostatic interaction is enhanced. 
PE brushes form increasing numbers of condensed layers (e.g., from two at $\chi=1.70$ to four at $\chi=1.30$) to increase the average separation distance between charges. The widths of these additional layers reflect the size of an electrostatic blob, approximately the Debye screening length $\kappa_D^{-1}$. 
For a given $\alpha$ and $\chi$, the maximum polymer density of each condensed layer is almost the same, which is determined by the maximum local charge density that can be maintained. The distance between two adjacent condensed layers is determined by balancing the electrostatic repulsion between the layers and the elastic energy due to chain stretching. 
Our theoretical prediction of these multi-layer morphologies is consistent with the experimental findings by Dura et al. and Randall et al. of the lamellae structure composed of alternating layers of ionomer Nafion and water at the interface between the substrate and the hydrated Nafion film \cite{dura_multilamellar_2009, randall_morphology_2024}.

As $\chi$ further decreases to $\chi=1.20$ (Fig.~\ref{fig:pha4Cd}), an additional swollen layer of lower polymer density is melted from the inner condensed layers and extends far into the solution. 
The coexistence between inner condensed layers and an outer swollen layer predicted by our theory is in good agreement with the experimental observations via cryo-TEM images of PSS-grafted particles by Samokhina et al. \cite{samokhina_binding_2007} 
This also explains the non-trivial neutron reflectivity spectra reported
by Reinhardt et al. in the study of neutral and charged brushes, where coexistent condensed and swollen layers were observed by fitting the spectra data.\cite{reinhardt_fine-tuning_2013, ballauff_phase_2016}. 
Similar coexisting morphologies were also observed in neurofilament-derived protein brushes at intermediate salt concentrations. Using the same continuous-space SCFT as used in this work, we have previously shown quantitative agreement with experimentally measured brush heights \cite{yokokura_effects_2024, ding_dissecting_2024}.
Finally, when the hydrophobicity becomes low, e.g. $\chi = 1.00$ (Fig.~\ref{fig:phaS}), PE brushes are fully swollen, recovering the scenario of the Alexander---de Gennes brush in the salted regime \cite{borisov_diagram_1994, rubinstein_polymer_2003}. Consistent with previous work \cite{israels_charged_1994, zhulina_theory_1995}, the brush height $H$ decreases as salt concentration $c_\pm^b$ increases, approximately following the scaling of $H \sim (c_\pm^b)^{-1/3}$ (see Fig.~S1 in the SI).

\begin{figure*}[ht!]
    \centering
    \includegraphics[width=0.8\textwidth]{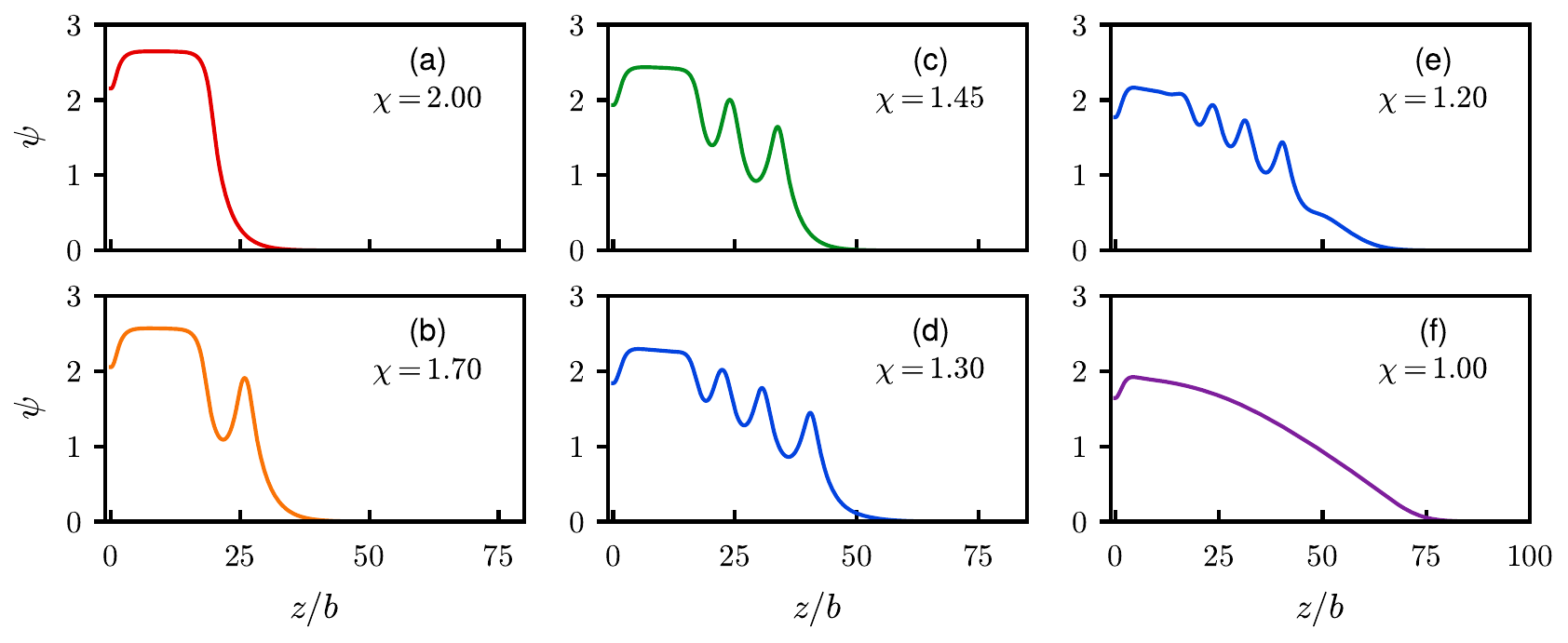} 
    \caption{Electrostatic potential profiles $\psi$ corresponding to the multi-layer morphologies shown in Fig.~\ref{fig:pha}.}
    \label{fig:el}
    {\phantomsubcaption{}\label{fig:psi1C} \phantomsubcaption{}\label{fig:psi2C} \phantomsubcaption{}\label{fig:psi3C} \phantomsubcaption{}\label{fig:psi4C} \phantomsubcaption{}\label{fig:psi4Cd} \phantomsubcaption{}\label{fig:psiS}}
\end{figure*}

Fig.~\ref{fig:el} plots representative electrostatic potential profiles $\psi$ corresponding to the multi-layer morphologies shown in Fig.~\ref{fig:pha}. The distributions of cations and anions are also shown in Fig.~S2 in the SI. The multi-peak features are clearly observed in the electrostatic potential profiles, similar to their corresponding polymer density distributions. $\psi$ exhibits peaks corresponding to each polymer-rich layer and decays in each solvent-rich layer. The peak value of $\psi$ for each polymer-rich layer decreases as the layers become further away from the grafting surface. Furthermore, as shown in Fig.~\ref{fig:psi4Cd}, the appearance of a ``shoulder'' at the outer edge of the $\psi$ profile signifies the existence of the additional swollen layer. Therefore, the multi-layer morphology of the PE brushes can be directly detected, if the average electrostatic potential in each layer can be measured experimentally.

\begin{figure}
    \centering
    \includegraphics[width=0.8\linewidth]{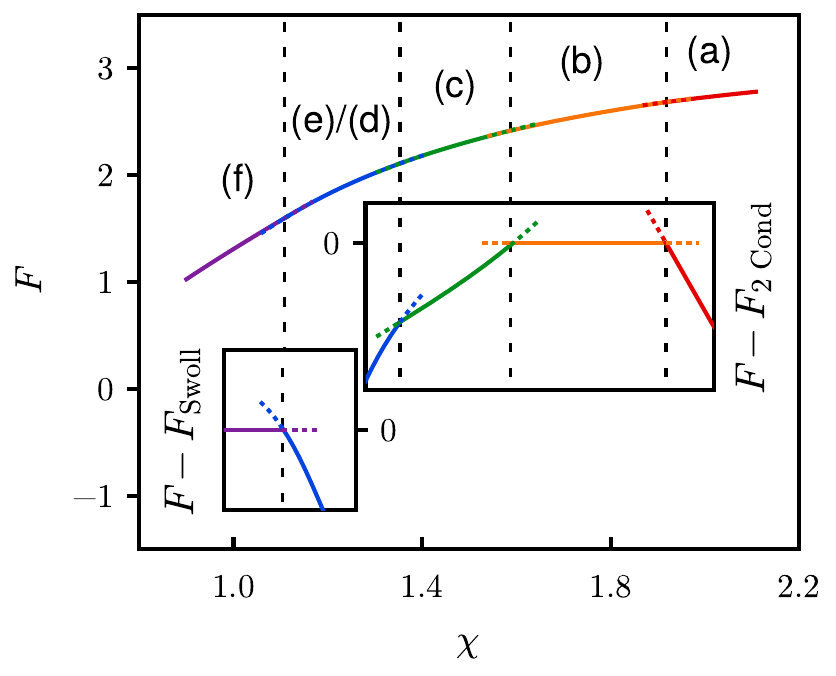} 
    \caption{Free energy (in excess to the reference bulk salt solution) as a function of the hydrophobicity $\chi$ for fixed $\alpha=0.4$. The labels (a)--(f) correspond to the representative multi-layer morphologies shown in Fig.~\ref{fig:pha}. The dotted lines indicate the metastable regimes. Vertical dashed lines indicate the locations of discontinuous transitions between morphologies. Insets show excess free energies with respect to (left) the fully swollen morphology and (right) the morphology with two condensed layers to highlight the energy crossings. }
    \label{fig:free}
\end{figure}

To elucidate the nature of the morphological transitions, Fig.~\ref{fig:free} plots the free energies of the multi-layer structures in their existence regimes as a function of the hydrophobicity $\chi$. 
Morphologies with more condensed layers become more stable as $\chi$ decreases, finally melting to a fully swollen brush. 
The crossings of the free energies as shown in the insets clearly indicate that the transitions between morphologies with consecutive numbers of condensed layers are discontinuous. The melting transition to the fully swollen brush is also discontinuous. 
The first-order nature of these transitions suggests a collective reorganization of all PE chains. 
On the other hand, it is interesting to note that the pre-swollen transition (around $\chi=1.25$) from the morphology with four condensed layers (Fig.~\ref{fig:pha4C}) to that with an additional swollen layer (Fig.~\ref{fig:pha4Cd}) is continuous. This continuous nature suggests a gradual reorganization of individual chains in the formation of the outer swollen layer.

\begin{figure}
    \centering
    \includegraphics[width=0.8\linewidth]{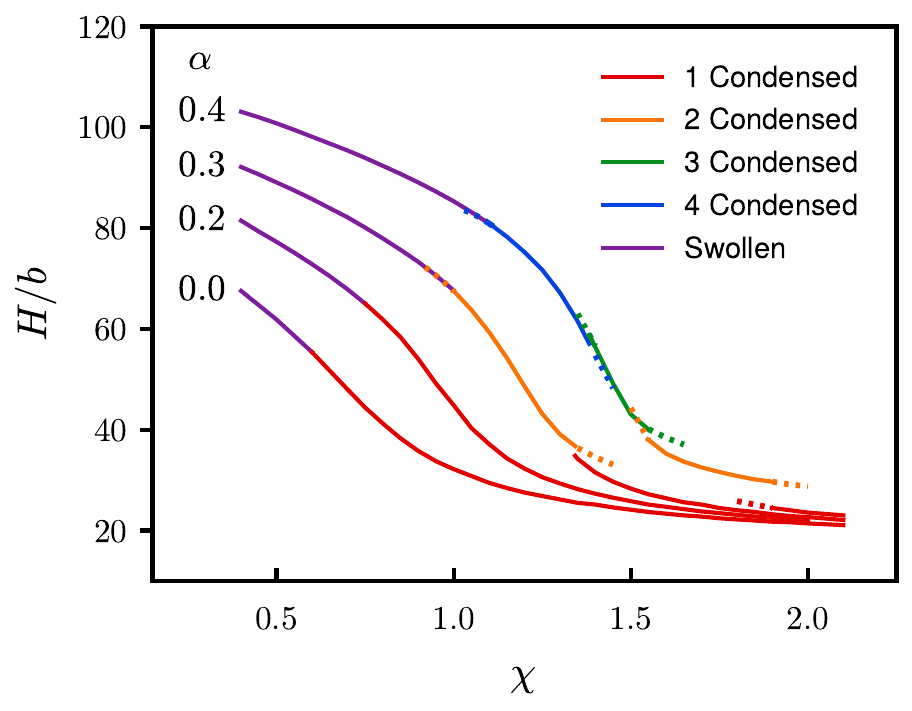}
    \caption{The effect of backbone charge density on the structural heterogeneity of PE brushes. The brush height $H$ (scaled by the Kuhn length $b$) is plotted as a function of $\chi$ at four different levels of $\alpha$. $H$ is defined as the distance at which the polymer density reduces to a threshold $\phi_p^* = 1\times10^{-6}$. Different morphologies are represented using different colored lines. The dotted lines indicate the metastable regimes.}
    \label{fig:heights}
\end{figure}

By systematically including the electrostatics into the polymeric self-consistent field theory, we are able to capture the interplay between the charge repulsion and hydrophobic attraction on the structural heterogeneity in PE brushes. 
Fig.~\ref{fig:heights} shows the evolution of multi-layer morphologies for different backbone charge densities $\alpha$. 
The morphology is characterized by the brush height $H$, a macroscopic measurable property which can be obtained experimentally using methods like ellipsometry and AFM. 
To facilitate the characterization of the multi-layer structures, $H$ in the current work is defined as the distance at which the polymer density reduces to a threshold $\phi_p^* = 1\times10^{-6}$. Another commonly adopted definition of brush height based on the Gibbs dividing surface  \cite{israelachvili_intermolecular_2011, wang_theory_2014} is not sensitive to structural heterogeneity. 

As shown in Fig.~\ref{fig:heights}, PE brushes exhibit richer morphological behaviors with more accessible multi-layer structures as $\alpha$ increases. 
Due to the absence of the long-range electrostatic repulsion, charge neutral brushes ($\alpha=0$) only exhibit a continuous transition from a fully condensed morphology to a fully swollen morphology. 
As $\alpha$ increases to 0.2, it is interesting to note that although an outer swollen layer emerges in the pre-swollen process, the transition remains continuous. This is consistent with early theoretical work invoking the assumption of local charge neutrality \cite{zhulina_structure_1992}. 
For larger backbone charge densities (e.g., $\alpha\geq 0.3$), morphologies with multiple condensed layers appear.
As $\alpha$ increases up to $0.4$, morphologies with two, three and four condensed layers are all accessible. 
If the brushes are in the salted regime \cite{borisov_diagram_1994}, a similar trend can also be expected as decreasing the bulk salt concentration $c_\pm^b$ (see Fig.~S1 in the SI).
The stronger electrostatic repulsion provides a larger driving force to overcome the energetic penalty associated with the increase in surface area when a new condensed layer is generated.  

As an initial attempt to study the multi-layer morphologies in PE brushes, we assume the same dielectric constants for polymers and solvent. However, as indicated by Eq.~\ref{SCF:main}, our theory is general for non-uniform dielectric environments. In many systems, polymers may have a much lower dielectric constant than solvent, which leads to a stronger screening effect on the electrostatic repulsion within the layers. This will shift the transitions between morphologies with consecutive numbers of layers to lower values of $\chi$, thus suppressing the formation of multi-layer morphologies (see Fig.~S3 in the SI). Therefore, lowering the dielectric constant of polymers has the similar effect as decreasing $\alpha$.
Furthermore, the current study is focused on dense brushes where $R_g^2\sigma \gg 1$. We assume that the formation of multi-layers via microphase segregation in the normal direction to the surface is favorable in the dense brush regime. We note that as the grafting density decreases, lateral microphase segregation in the form of cylindrical aggregates, pinned micelles, etc. may also occur. We expect that laterally segregated morphologies will be energetically favorable in the regime of lower grafting density \cite{carrillo_morphologies_2009}. 

\subsection{Molecular Origin of Multi-layer Morphologies}

\begin{figure}[ht!]
    \centering
    \includegraphics[width=1.0\linewidth]{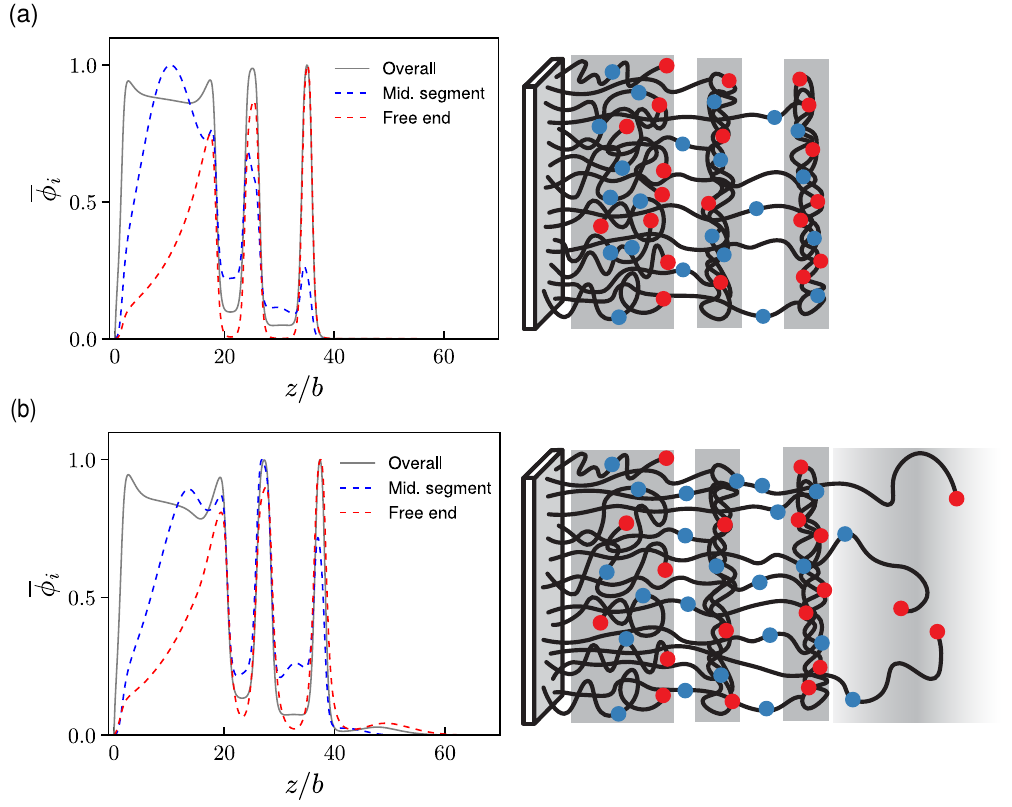}
    \caption{Density distribution of specific segments in (a) brushes with three condensed layers ($\alpha=0.4$ and $\chi=1.40$) and (b) pre-swollen brushes with three condensed layers and an outer swollen layer ($\alpha=0.4$ and $\chi=1.30$). In the left column, density distributions $\overline{\phi}_i$ for the overall polymer, middle segment and free end are normalized by their respective maxima. The right column provides the corresponding schematic.  
    The middle segments are marked by blue circles while the free ends are marked by red circles. The gray, shaded regions illustrate the overall polymer density of each layer. }
    {\phantomsubcaption{}\label{fig:dist3C} \phantomsubcaption{}\label{fig:dist3Cd}}
\end{figure}

Our SCFT can also provide the distributions of individual segments, which help us to understand the arrangement of the PE chains in multi-layer morphologies at the molecular level. 
Fig.~\ref{fig:dist3C} plots the density distributions corresponding to the overall polymer, middle segment and free end for a brush containing three condensed layers ($\alpha=0.4$ and $\chi=1.40$). Each distribution $\overline{\phi}_i$ is normalized by its corresponding maximum to facilitate the comparison. 
Notably, the middle segments and free ends are both distributed in all three polymer-rich layers. While the middle segments are more likely to be found in the inner layers, the free ends prefer to be distributed in layers further away from the substrate. 
Furthermore, the free ends are almost absent in the polymer-poor regions between two adjacent condensed layers, in contrast to the middle segments. 
These features indicate that multi-layer brushes are composed of separate subpopulations of PE chains, as shown in the schematic of Fig.~\ref{fig:dist3C}. 
Individual chains choose to segregate into one of the condensed layers. Starting from the grafted end, a given chain may remain adjacent to the substrate
to form the innermost layer or choose to stretch further to form one of the outer layers. At
the middle segment, the chain either has arrived at its chosen layer or is still on its way. Finally, the free end is always in its chosen condensed layer. 

We emphasize that the microphase segregation in multi-layer PE brushes is significantly different from that in the pearl-necklace structure of single PE chains in poor solvents \cite{dobrynin_cascade_1996, dobrynin_theory_2005, liu_variational_2024}. 
The multi-layer brushes are not simply a parallel alignment of pearl-necklaces. The segregation is not formed by different portions of the same chain, but by different subpopulations of all chains. 
This arrangement of different polymers in different condensed layers optimizes the total elastic energy caused by chain deformation. 

To further uncover the nature of the pre-swollen morphology, Fig.~\ref{fig:dist3Cd} illustrates the corresponding distributions for brushes containing three condensed layers and one outer swollen layer ($\alpha=0.4$ and $\chi=1.30$). 
The features of the condensed layers are generally consistent with those in Fig.~\ref{fig:dist3C}. 
On the other hand, the outer swollen layer has a rather wide distribution and is composed of a much larger fraction of free ends than middle segments. This signifies that the chains in the swollen layer are fully stretched as shown in the schematic of Fig.~\ref{fig:dist3C}. 
In a pre-swollen brush, the hydrophobic attraction is not sufficient to confine all the chains into condensed layers such that the electrostatic repulsion forces a subpopulation of chains to melt. 
As $\chi$ further decreases (or equivalently $\alpha$ increases), the electrostatic repulsion dominates, making microphase segregation unfavorable.  All PE chains become fully stretched and the brush thus adopts the swollen morphology. 
Therefore, the pre-swollen process is a prerequisite for multi-layer structures to fully melt.

\subsection*{Scattering Behavior of Multi-layer Morphologies}
Commonly used characterization techniques like ellipsometry and AFM can only provide the overall brush height, which is not sufficient to effectively reflect the microstructure of multi-layer morphologies in PE brushes. On the other hand, reflectivity is a scattering technique that detects the detailed structure of buried layers by utilizing the interference of X-rays or neutrons. However, extracting density profiles directly from reflectivity spectra is quite challenging because the scattering intensity is measured in reciprocal space \cite{skoda_recent_2019}. The reflected intensity $R$ of an X-ray beam is related to the gradient of the electron density distribution $\mathrm{d}\rho_e(z)/\mathrm{dz}$ \cite{russell_x-ray_1990}:
\begin{equation}\label{eq:master}
R(Q_z)/R_F = \left| \frac{1}{\rho_e^{\infty}}\int \textrm{dz}\ \frac{\mathrm{d}\rho_e(\mathrm{z})}{\mathrm{dz}} \exp{(iQ_z \mathrm{z})} \right| ^2~,
\end{equation}
where $R_F$ is the Fresnel reflectivity of an ideal, sharp surface. $Q_z$ is the specular part of the momentum transfer vector and $\rho_e^{\infty}$ is the electron density of the bulk media far away from the substrate. The local electron density $\rho_e(\mathrm{z})$ can be calculated from the distribution of each species $\alpha$ based on the linear combination $\rho_e(\mathrm{z}) = \sum_{\alpha}\rho_{e,\alpha}\phi_\alpha(\mathrm{z})$. The species $\alpha$ considered here are water, polymer, and the \ch{SiO2} substrate, with the corresponding electron densities 0.33, 0.58, 0.70 $e\ \mathrm{\angstrom}^{-3}$, respectively. The density distributions $\phi_\alpha(\mathrm{z})$ obtained by our SCFT can thus be used straightforwardly to generate the reflectivity spectra $R(Q_z)$.

\begin{figure}
    \centering
    \includegraphics[width=1.00\linewidth]{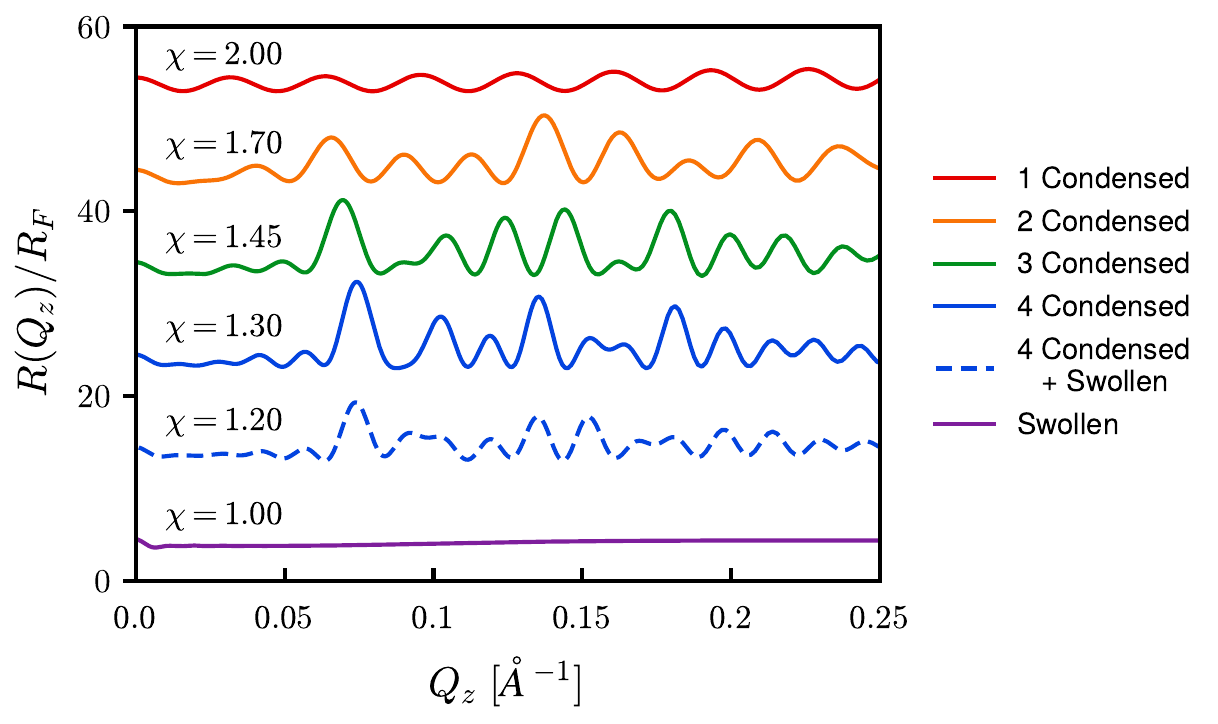}
    \caption{Reflectivity spectra in multi-layer PE brushes. The reflected intensity $R$ (normalized by Fresnel reflectivity $R_F$) is plotted against the specular part of the momentum transfer vector $Q_z$. Spectra are shifted vertically by 10 equiv for clarity. The corresponding polymer density distributions can be found in Fig.~\ref{fig:pha}.}
    \label{fig:refl}
\end{figure}

Fig.~\ref{fig:refl} shows the evolution of the reflectivity spectra for different multi-layer structures formed in PE brushes. 
The key features of the reflectivity spectra are the oscillation period, which is related to layer thickness, and the oscillation amplitude, which is related to the sharpness of the polymer-solvent interface. 
The spectra exhibit uniform oscillations for morphologies with a single condensed layer ($\chi = 2.00$). 
As the number of condensed layers further increases to two ($\chi = 1.70$), three ($\chi = 1.45$) and four ($\chi = 1.30$), the number of polymer-solvent interfaces that can reflect X-rays also increases. 
This reduces the period of the oscillations and increases the sharpness of the peaks.
In addition, the irregular distances between these interfaces lead to interference between their reflected X-rays, thus generating complex, non-uniform spectra. 
The oscillations do not maintain a constant amplitude and period. 
Furthermore, with the increase of structural heterogeneity, more energy is required for the X-rays to reach the substrate. This increases the momentum $Q_z$ required to obtain appreciable amplitude in the reflectivity spectra, and thus shifts the total spectra to higher $Q_z$. 
While keeping the same multi-layer morphology, adding more salt increases the screening effect which makes the layers more compact. The resulting spectra also shifts to higher $Q_z$ (see Fig.~S4 in the SI). 

The addition of the outer diffuse layer in the pre-swollen morphology ($\chi=1.20$) adds slightly more interference to the spectra and reduces the amplitude of the oscillations. 
Finally, for fully swollen brushes ($\chi=1.00$), the polymer-solvent interface is diffuse and the reflectivity spectra is essentially flat and nearly featureless. 
Therefore, the features of the non-uniform oscillating spectra signify the multi-layer morphologies of PE brushes. 
The key features of the spectra predicted by our theory are also consistent with those observed in experiments. For instance, Dura et al. found multilayer morphologies in Nafion films which show oscillations in the spectra with a high degree of interference for $Q_z > 0.06~\angstrom^{-1}$ \cite{dura_multilamellar_2009}. This is close to the onset of non-uniform oscillations found in our theoretical work. In addition, as the number of condensed layers increases, the period of the oscillations were found to increase \cite{randall_morphology_2024}, which is also in agreement with the features shown in Fig.~\ref{fig:refl}.

Our SCFT calculations bridge the experimentally measurable reflectivity spectra to the microstructure of multi-layer brushes. For an experimentally measured spectrum, a series of candidate density profiles can be generated by our SCFT, which can then be converted to corresponding reflectivity spectra. The one with the best agreement with the experimental result can be used to identify the brush morphology.

%% file: 4_conclusions.tex
\section*{Conclusions}
In this work, we apply a continuous-space self-consistent field theory to study the structural heterogeneity of PE brushes with high grafting densities. 
We predict that the competition between electrostatic repulsion and hydrophobic attraction induces microphase segregation, forming a series of multi-layer morphologies with alternating polymer-rich and polymer-poor domains. 
As the relative importance of electrostatic repulsion increases, morphologies with larger numbers of condensed layers become accessible, which eventually melt into fully swollen brushes via a pre-swollen stage. 
We find that the transitions between morphologies with consecutive numbers of condensed layers as well as the melting transition to the fully swollen brush are all discontinuous. 
Through analyzing the distribution of specific chain segments, we elucidate that the segregated layers are not formed by different portions of the same chain, but by different subpopulations of all chains. This is significantly different from the scenario of the pearl-necklace structure formed by a single PE chain in poor solvents. 
Furthermore, our theory provides an effective tool to bridge experimentally measurable reflectivity spectra to the microstructure of multi-layer brushes. We show that the oscillation period and amplitude in the spectra are very sensitive to the number of layers and the sharpness of the polymer-solvent interfaces. 
The multi-layer morphology predicted by our theory is in good agreement with the lamellae structure experimentally observed at the interface between hydrated Nafion film and substrate.

In the current work, we use a simple system of PE brushes composed of homopolymers with uniform backbone charge density to illustrate the effect of competing interactions on the structural heterogeneity. Our theory can be straightforwardly generalized to brushes with various chain architectures, copolymer compositions and charge patterns. It can also be easily implemented to study protein brushes with specific amino acid sequences. Similar multi-layer morphologies with separate chain populations in each layer have been found in our previous studies on intrinsically disordered protein brushes \cite{yokokura_effects_2024,ding_dissecting_2024}.

Here, we focus on the multi-layer morphologies formed by microphase segregation of PE brushes in the normal direction. 
However, solving SCFT in high dimensions for brushes with varying grafting densities may lead to segregation in other directions with rich surface patterns \cite{carrillo_morphologies_2009}. We leave the exploration of the full phase diagram for a future work.
Furthermore, current applications of PE brushes largely rely on their steric and electrostatic repulsion to protect the surface underneath. 
As revealed in this work, designing structurally heterogeneous PE brushes by tuning their electrostatic and hydrophobic interactions could regulate surface properties at a higher level.

%% file: 5_SI.tex
\onecolumn
\subsection*{\Large{Supplemental Information}}
\bigskip 
\renewcommand{\theequation}{S\arabic{equation}}
\setcounter{equation}{0}
\renewcommand{\thefigure}{S\arabic{figure}}
\setcounter{figure}{0}
\renewcommand{\thetable}{S\arabic{table}}
\setcounter{table}{0}

\subsection{I. Effect of Bulk Salt Concentration on the Brush Morphology}

\begin{figure}[ht!]
    \centering
    \includegraphics[width=0.65\linewidth]{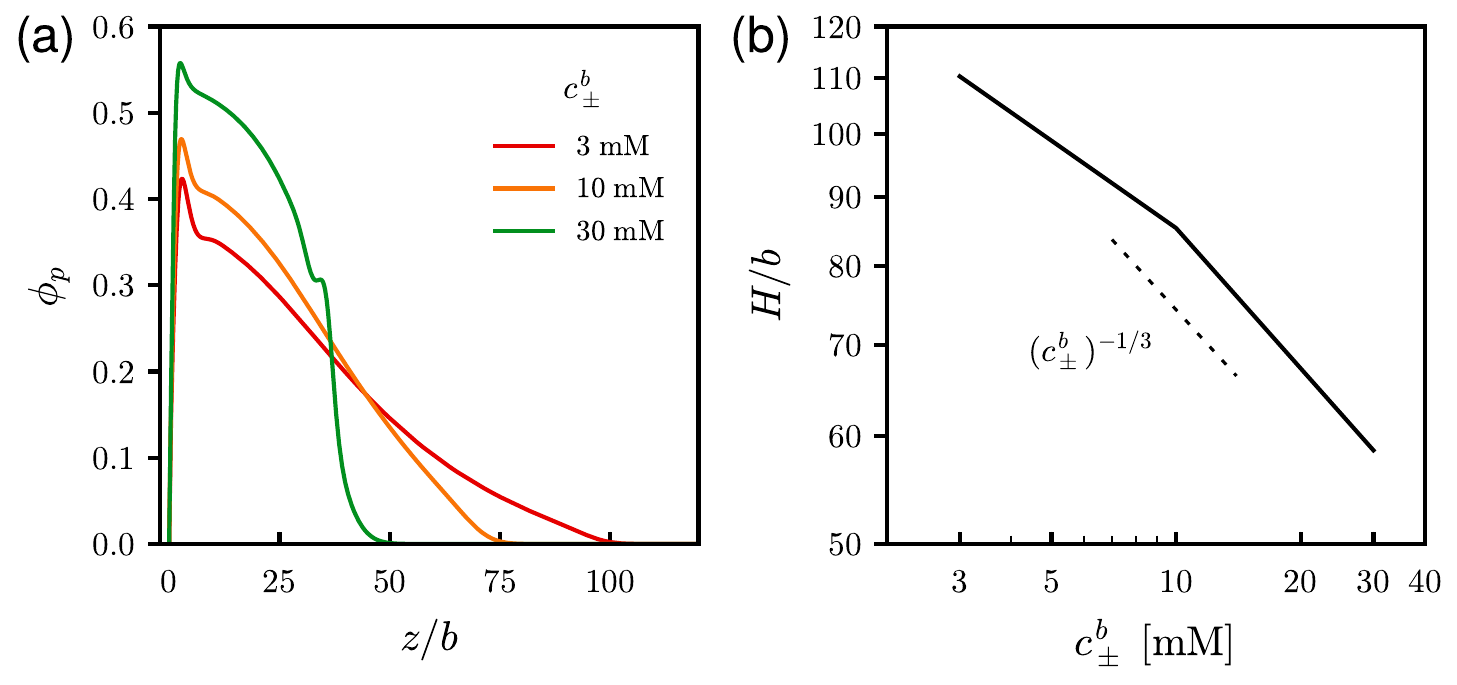}
    \caption{Effect of bulk salt concentration on (a) the density profile and (b) brush height $H$ of swollen PE brushes with $\alpha=0.4$ and $\chi=1.00$. The dotted line in (b) represents the scaling result for the salted regime, $H\sim(c_\pm^b)^{-1/3}$.}
    \label{fig:salted}
    {\phantomsubcaption\label{fig:salted-a}\phantomsubcaption\label{fig:salted-b}}
\end{figure}

We examine the dependence of the brush morphology on the bulk salt concentration. We plot the polymer density profile in Fig.~\ref{fig:salted-a} and brush height in Fig.~\ref{fig:salted-b}. We find brushes shrink as salt concentration increases. The height almost follows the scaling of $H\sim (c_\pm^b)^{-1/3}$, suggesting that the brushes studied here are in the salted regime. Therefore, adding salt indeed has an impact on the multi-layer morphology. Increasing the salt concentration enhances the screening effect and thus reduces the electrostatic repulsion within the brushes, sharing a similar effect as reducing the backbone charge density $\alpha$. We expect that as the salt concentration increases, the multi-layer structures become less accessible.

\newpage
\subsection{II. Electric Double Layer Structure within Multi-layer Morphologies}

\begin{figure}[ht!]
    \centering
    \includegraphics[width=0.9\linewidth]{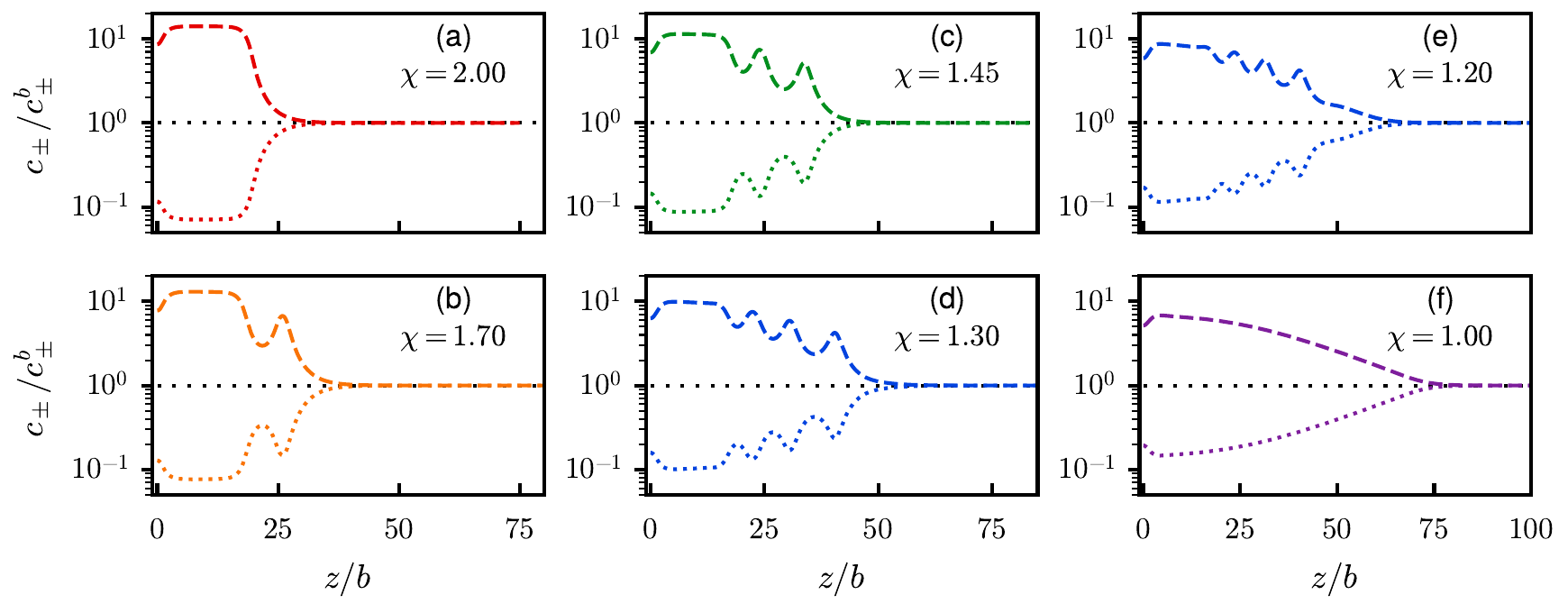}
    \caption{Distributions of cations (dotted lines) and anions (dashed lines) in different multi-layer morphologies. The concentration of ions $c_\pm$ (normalized by the corresponding bulk concentration $c_\pm^b$) is plotted in the $z$-direction normal to the substrate. The parameters used in (a) to (f) correspond to those in Fig.~2(a) to (f) in the main text.}
    \label{fig:anca}
\end{figure}

\newpage
\subsection{III. Effect of Polymer Dielectric Constant on the Stability of Multi-layer Morphologies}
\begin{figure}[ht!]
    \centering
    \includegraphics[width=0.50\linewidth]{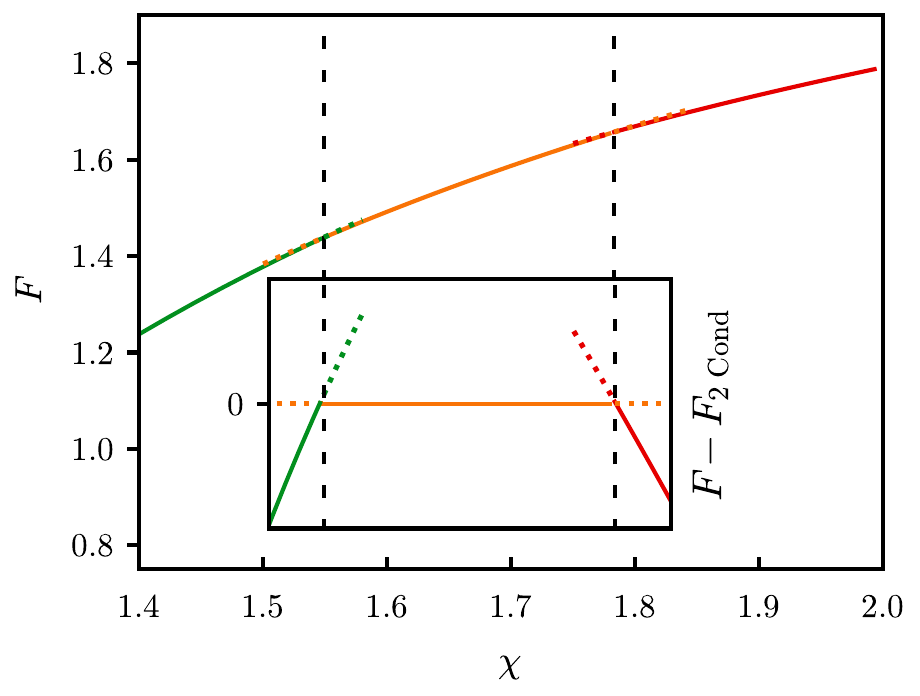}
    \caption{Free energy (in excess to the reference bulk salt solution) as a function of the hydrophobicity $\chi$ for $\epsilon_{p}=10$ and $\epsilon_{s}=80$ with fixed $\alpha=0.4$. The dotted lines indicate the metastable regimes. Vertical dashed lines indicate the locations of discontinuous transitions between morphologies. Inset shows excess free energies with respect to the morphology with two condensed layers to highlight the energy crossings. The local dielectric constant is calculated from the linear combination of that of the polymer and solvent, $\epsilon_r(\br) = \epsilon_{p}\phi_p(\br)+\epsilon_{s}\phi_s(\br)$.}
    \label{fig:f_eps}
\end{figure}

Our theory can also include the non-uniform dielectric permittivity. Here, we performed calculations using polymers with lower dielectric constant $\epsilon_{p} = 10$. The local dielectric constant is taken to be the linear combination of that of the polymer and the solvent, $\epsilon_r(\br) = \epsilon_{p}\phi_p(\br) + \epsilon_{s}\phi_s(\br)$. Fig.~\ref{fig:f_eps} shows the free energy as a function of $\chi$ for the multi-layer morphologies calculated at $\epsilon_p = 10$. Compared to Fig.~4 in the main text, the transition points between consecutive layers shift to lower values of $\chi$. In general, lowering $\epsilon_p$ has the same effect as lowering the backbone charge density $\alpha$.  The reason is that lowering $\epsilon_p$ increases the screening effect on the electrostatic repulsions within the layers, which makes the hydrophobic attraction more dominant in the competition. This effect suppresses multi-layer formation.

\newpage
\subsection{IV. Effect of Bulk Salt Concentration on the Reflectivity Spectra}

\begin{figure}[ht!]
    \centering
    \includegraphics[width=0.75\linewidth]{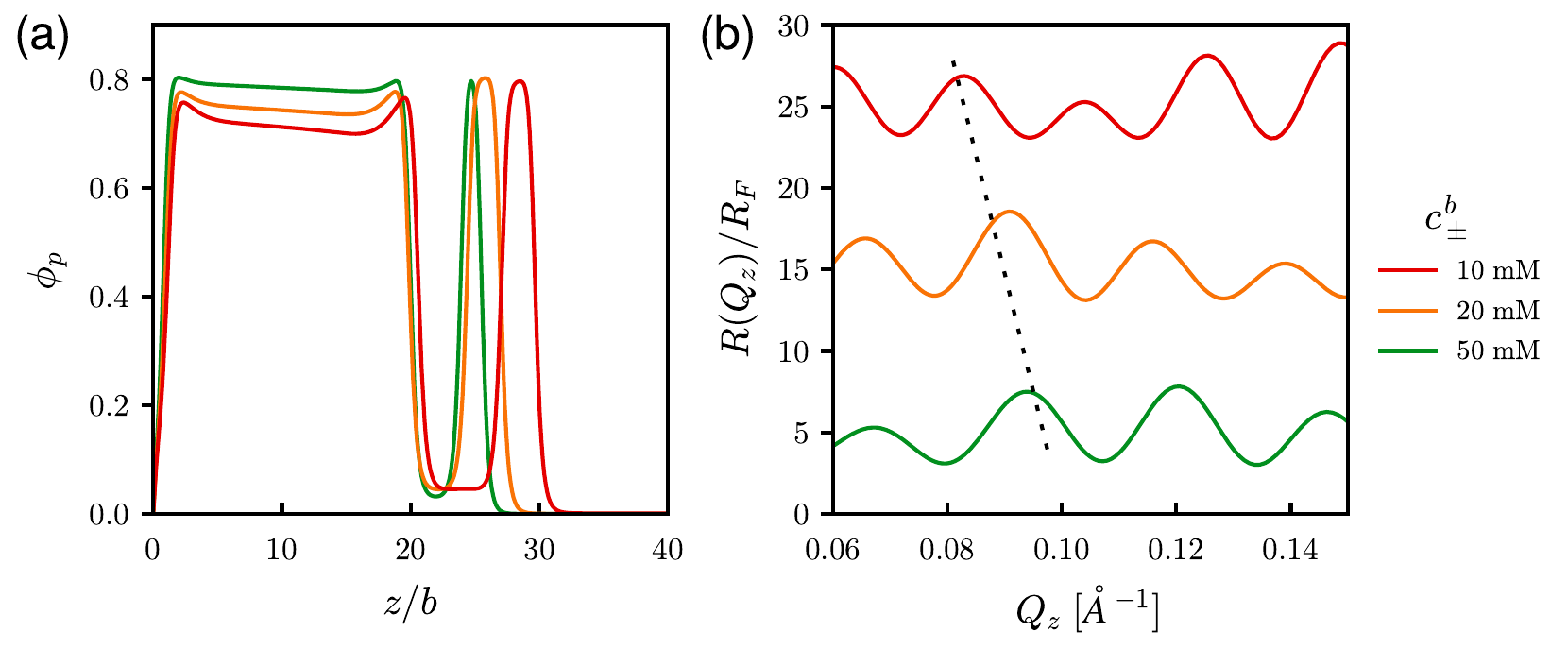}
    \caption{Evolution of reflectivity spectra at three different salt concentrations for PE brushes with $\alpha=0.4$ and $\chi=1.45$. (a) Density distributions and (b) corresponding reflectivity spectra for PE brushes with two condensed layers. For clarity, the curves are offset vertically by 10 equiv as $c_\pm^b$ increases. The dotted guideline illustrates the shift of the spectra to higher $Q_z$.}
    \label{fig:ref-cs}
\end{figure}

We examine the salt effect on the spectra. Fig.~\ref{fig:ref-cs} shows the effect of salt concentration on the density profile and reflectivity spectra of the two-layered morphology. Increasing the salt concentration increases the screening effect and thus makes the layers more compact. In the spectra, the peaks shift to higher $Q_z$.

